# Determining the True Optical Gap in a High-Performance Organic Photovoltaic Polymer Using Single-Molecule Spectroscopy

Gordon J. Hedley*, Florian Steiner, Jan Vogelsang and John M. Lupton

Institut für Experimentelle und Angewandte Physik, Universität Regensburg, D-93040, Regensburg, Germany

*E-Mail: Gordon.Hedley@physik.uni-regensburg.de

**Abstract**

Low-gap conjugated polymers have enabled an impressive increase in the efficiencies of organic solar cells, primarily due to their red absorption which allows harvesting of that part of the solar spectrum. Here, we report that the true optical gap of one prototypical material, PTB7, is in fact at significantly higher energy than has previously been reported, indicating that the red absorption utilized in these materials in solar cells is entirely due to chain aggregation. Using single-molecule spectroscopy we find that PL from isolated nanoscale aggregates consists of multiple independently emitting chromophores. At the single-molecule level, however, straight single chains with a high degree of emission polarization are observed. The PL is found to be ~0.4 eV higher in energy, with a longer lifetime than the red aggregates, and is attributed to single chromophores. Our findings indicate that the impressive light-harvesting abilities of PTB7 in the red spectral region arises solely from chain aggregation.

**Table of Contents Graphic**

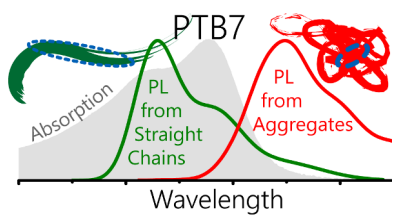

The photoelectric effect was first observed by Edmond Becquerel in 1839 with silver chloride.[1] In such a material light absorption occurs in the blue and ultraviolet spectral region. The solar spectrum, in contrast, peaks in the green and extends well into the infrared, thus when gathering sunlight to convert it into electrical energy, absorbing as much as possible at all wavelengths, from green to red, is crucial. In organic photovoltaics a conjugated polymer is the primary absorbing material, and sustained development over the last 20 years has seen power conversion efficiencies increase from ~2% to over 12%.[2-3] One chief driver of this increase in efficiency is the synthesis of conjugated polymers that absorb in the red and near-infrared region of the solar spectrum, converting previously lost photons at these wavelengths into electrical energy.[4-5] The family of polymers based around thienothiophene and benzodithiophene have become a benchmark for this class of red absorbing materials, achieving power conversion efficiencies of over than 10%.[6-7] Here, we focus on one of the best-performing polymers from this class, PTB7[8-9] (chemical structure shown in the inset of Figure 1), and find that its apparent red absorption is in fact entirely due to aggregation. When straight, single chains of PTB7 are formed, the actual fundamental electronic gap of this material is over 0.4 eV higher than is observed in ensemble absorption. As such, it is clear with this class of materials that one has to consider the aggregation of polymer chains as being integral to the impressive photovoltaic performance.

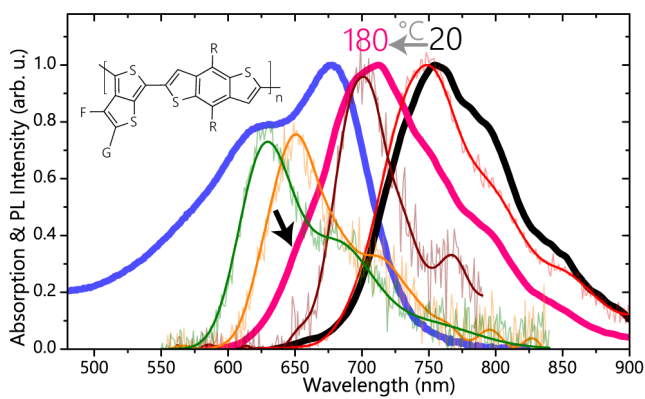

**Figure 1:** Absorption and photoluminescence (PL) spectra of PTB7 ensemble solutions in toluene (thick solid lines) at 20°C (black line) and 180°C (pink line). A shoulder at 650 nm is seen in the latter, indicated by an arrow. Four different single-molecule PL spectra are also shown (thin solid lines, with an average superimposed over the raw data), indicating that some single molecules match the ensemble PL at 20°C, while some match the PL from solutions at 180°C. Some single-molecule PL spectra lie entirely within the ensemble absorption spectrum. The chemical structure of PTB7 is given in the inset, with R = 2-ethylhexyloxy and G = 2-ethylhexylcarbonyl.

To begin our investigations on aggregation in PTB7 we examine its photoluminescence (PL) in solution at different temperatures, working at very low concentrations on a confocal microscope. The thick-lined curves in Figure 1 show such ensemble spectra excited at 488 nm. At room temperature the PL spectrum peaks at 750 nm, while at 180°C it shifts substantially to the blue, with the PL peak at 700 nm and a shoulder at 650 nm. Spectra for intermediate temperatures along with the return-cooled room-temperature solution showing no degradation due to heating are presented in *Supporting Information* (SI). The clear blue-shift with increasing temperature is consistent with deaggregation of polymer chains[10-11] rather than the presence of oligomers[12] as the entire spectrum blue-shifts, and, as shown in SI, becomes ~3 times more emissive, which is indicative of reduced aggregation quenching. Absorption spectra at elevated temperatures have previously been reported[10] for a polymer very similar to PTB7, and also show a blue-shift with increasing temperature. Specifically, they show a loss of the main absorption peak at 690 nm, leading to a single peak at 620 nm. Another possible explanation of the blue-shift is that a barrier to rotation of polymer chain segments is overcome at elevated temperatures, enabling the chain to split into smaller, bluer, segments when it is heated. To truly understand the nature of these non-aggregated chains is, however, difficult in solution spectroscopy, as an ensemble is measured and experiments require working at 180°C or ideally higher, which is not particularly practicable. We thus choose to utilize single-molecule spectroscopy to examine PTB7 chains in the solid state that are isolated from each other when embedded in an inert host matrix; this allows us to break down the ensemble into its constituent parts and to study them in greater detail.

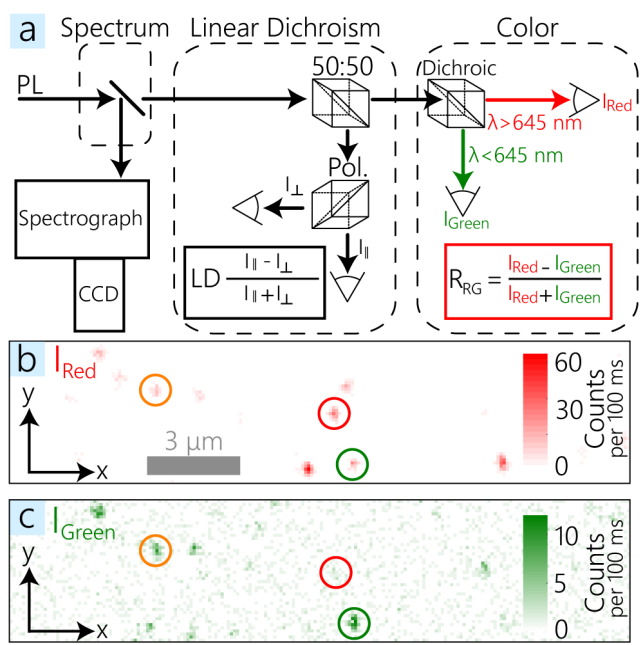

**Figure 2: a)** Schematic of the microscope detection optics, with the option of measuring PL spectra or measuring the linear dichroism (LD) and red-green color ratio ($R_{RG}$), as defined in the inset, on 4 avalanche photodiodes. **b)** Confocal scan image of an area of ~20x5 µm² showing the signal detected only on the red APD, with the same area imaged in **(c)** on the green APD, enabling $R_{RG}$ and the individual spot colors to be determined. Three exemplar spots are circled, red for an $R_{RG}$ ~ 1, orange for an $R_{RG}$ ~ 0.3 and green for an $R_{RG}$ ~ 0, with the associated colored single-molecule PL spectra in Figure 1 approximately corresponding to these $R_{RG}$ values.

Single molecule spectroscopy is performed with a confocal microscope with 488 nm pulsed laser diode excitation at 80 MHz repetition rate. As depicted in Figure 2a, we utilize different detection methods depending upon the experiment, with PL from the sample being directed to 4 avalanche photodiodes (APDs), or to a spectrograph-coupled CCD detector to record PL spectra. Full details of the setup are provided in the Experimental Methods. Samples of PTB7 at single-molecule concentrations are prepared by doping the material at ever lower concentrations in a poly(methyl-methacrylate) (PMMA) host until a regime is reached where well-isolated diffraction-limited spots are observed, as shown in the confocal scan images in Figure 2b,c. By observing the PL spectra from these isolated spots we find that a broad range of emission wavelengths are detected, shown in Figure 1 as the thin solid line spectra. Single-molecule PL spectra range from ones that match well with the ensemble PL, through to ones that are substantially blue-shifted and lie entirely within the ensemble *absorption* spectrum of PTB7. We note that these blue-shifted single-molecule spectra match well with the ensemble solution PL spectrum at 180°C, both the 700 nm peak and also the 650 nm shoulder, indicating that the room-temperature single-molecule chain conformation can indeed be unaggregated. However, single chains can show even bluer PL down to ~600 nm, indicating that further reduction in aggregation is possible on the single-chain level compared to the high-temperature ensemble solutions.

In order to examine many spots and monitor photophysical properties (lifetime, red-green color ratio, and linear dichroism (LD)) we remove the spectrograph:CCD unit and instead route the PL from each spot on the film through a 50:50 beam splitter. The first 50% of the PL is further divided with a polarizing beam splitter into PL polarized parallel and perpendicular to the laser field and detected with two APDs, which allows us to calculate the emission LD as defined in Figure 2a. The other 50% is passed onto a dichroic beam splitter centered at a wavelength of 645 nm and the respective wavelengths are then detected on two APDs, denoted here as $I_{Red}$ and $I_{Green}$, thereby enabling us to sort emission spots based on the ratio $R_{RG}$ of red to green emission as defined in Figure 2a. This ratio determines the color of the spot without having to record the entire PL spectrum. We can simultaneously record confocal scan images from both red and green APDs, as shown in Figure 2b and c. Spots dominated by PL in the red channel are most common (an example is highlighted with a red circle in the maps), with $R_{RG}$ values ~1. There are some spots that show a slightly more uniform distribution between the two detectors, with $R_{RG}$ ~ 0.3 (highlighted with an orange circle in the maps). And finally there are a few spots that show an approximately even split between the detectors with $R_{RG}$ ~ 0 (highlighted with a green circle in the maps). The use of confocal images gives us an overall sense of the percentages of spots that are red, orange or green in our above definition of the approximate $R_{RG}$ regions – and we find that out of >1000 single spots ~10% of spots are green and orange (i.e. -0.3 < $R_{RG}$ < 0.6), while ~90% are red (0.85 < $R_{RG}$ < 1). These percentages change very little even if we go to very low concentrations of PTB7 in PMMA, as shown in the Supporting Information where we decrease the overall concentration by a factor of 20 and still approximately 10% of the observed emission is within the range -0.3 < $R_{RG}$ < 0.6. We conclude that PTB7 is a polymer which aggregates very easily, and it is very difficult to break up.

Using this configuration, we can obtain the dynamics of the linear dichroism of each spot along with its $R_{RG}$ value, PL intensity and PL lifetime using time-correlated single-photon counting. Each of these dynamics for single red, orange and green spots as marked in Figure 2 are shown in SI. Over the 30 seconds that we measure the PL we observe no significant variation in the PL intensity or in the values of $R_{RG}$ or LD for each of the three chosen spots: the single PTB7 chains are remarkably photostable compared to other conjugated polymers. The measured time-resolved PL decays for each of the three spots are also shown in the SI. All three fit well to mono-exponential decays, with the red spot having the fastest decay with a lifetime of 0.66 ns, while the orange and green spots are slower with lifetimes of 0.82 and 0.85 ns, respectively.

The overall trend that we highlight from the single exemplar spots of each color is solidified when we look at many single chains to build a bigger statistical picture, as shown in Figure 3, in which approximately 1000 spots are analyzed. The trends found are that the orange and green spots are much less emissive than the red spots (Figure 3a), that the orange and green spot PL lifetimes are longer than the red ones (Figure 3b) and that the LD of the orange and green spots is much higher than that of the red spots (Figure 3d and e). Examples of two PL decays are shown in the inset in Figure 3b. Representative PL spectra are shown in Figure 3c to illustrate the continuum of optical gaps, with colors coded to the approximate $R_{RG}$ values. The blue PL spectrum shown in Figure 3c is an extrapolation of the anticipated PL spectrum given the lowest recorded $R_{RG}$ values of -0.3. Since measurements of spectra are more cumbersome than measurements of $R_{RG}$, we did not succeed in actually measuring such a strongly blue-shifted PL spectrum. An $R_{RG}$ value of -0.3 implies a PL peak of ~600 nm, representing the shortest wavelength and thus largest optical gap for PTB7. The method for extrapolating such a spectrum from the $R_{RG}$ value is detailed in the SI. The question remaining is how many chromophore units on the polymer chain contribute to the different emission colors?

In order to determine the nature of the emitting species we measured the photon antibunching[13-16] by using a Hanbury Brown-Twiss configuration[17-19] with the microscope, employing a 50:50 beam splitter to direct the PL onto two avalanche photodiodes in order to examine the second-order cross-correlation between the two signals. The photon correlations are measured for red spots (Figure 3f) and green spots (Figure 3g) and show a small antibunching dip for the red spots, but a much larger one for the green spots. The dashed lines indicate the statistical limits that one would expect[20] for one and two emitting chromophores given the measured signal and noise levels, indicating that the green spots arise from a single emitter, while clearly the red-spot PL originates from many independent units (>5).

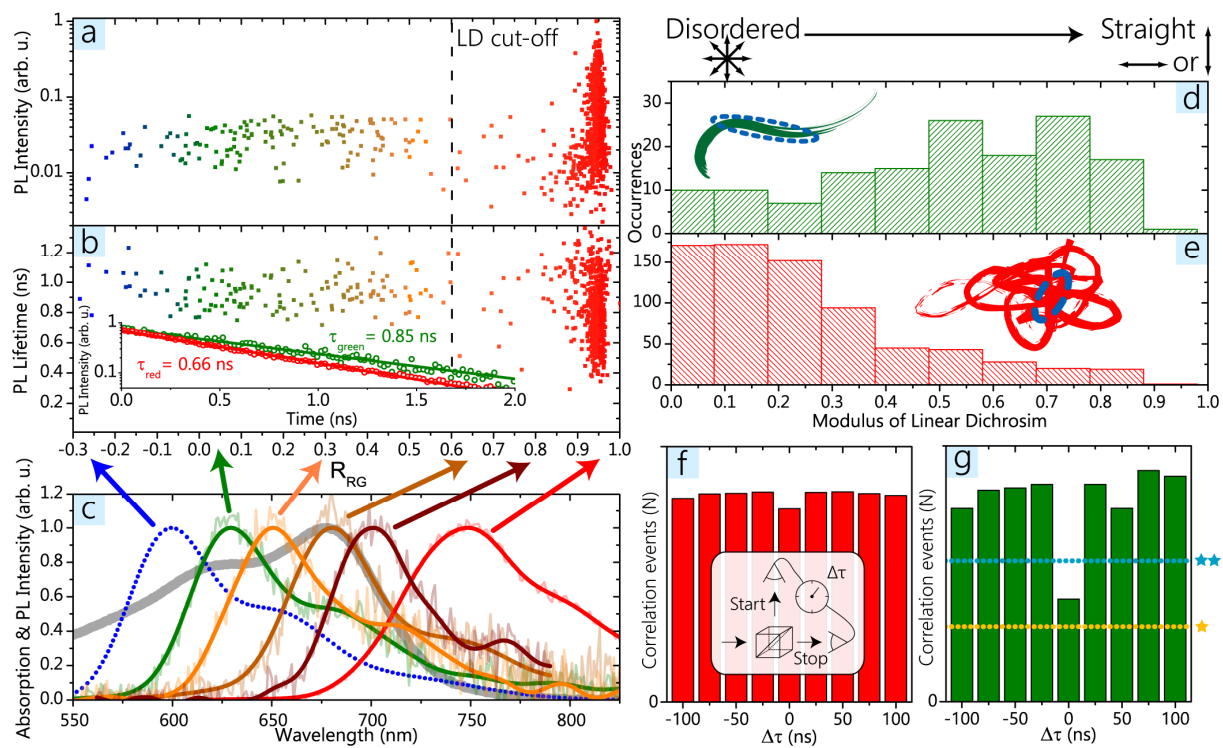

**Figure 3:** Intensity of single PL spots **(a)** and PL lifetimes **(b)** versus the measured $R_{RG}$ for approximately 1000 spots of PTB7. Red spots are substantially brighter and have shorter lifetimes. Exemplar single-exponential PL decays are shown in the inset of panel b. **c)** PL spectra of a selection of measured spots and the calculated $R_{RG}$ (indicated by arrows to the axis above). The blue dotted line is an inferred PL spectrum for the bluest measured $R_{RG}$ values, as described in the text. The ensemble absorption spectrum is shown as a thick gray line for reference. **d, e)** Histogram of linear dichroism for the same ~1000 spots, with the data in panel d having $-0.3 < R_{RG} < 0.6$, i.e. green and orange fluorescent spots showing high LD, and in panel e with $0.85 < R_{RG} < 1$, i.e. red spots showing low LD. Suggested polymer conformations are sketched in the inset for each case. **f, g)** Photon statistics for 59 red (f) and 19 green (g) spots as defined above, with the experimental schematic shown in the inset. A small antibunching dip (~5%) at time zero is observed in the red aggregates consistent with multi-chromophoric emission, contrasting with the much larger dip (~50%) observed in the green straight chains, consistent with a single emitting chromophore. The anticipated signal levels are shown as dashed lines for 1 (*) and 2 (**) emitting chromophores, calculated from the observed signal-to-noise ratio as described in the text.

The measured single-molecule results presented here for PTB7 allow us to build and discuss a picture of the likely nature of the emitting chromophores for this polymer. The brightness of the red spots combined with their short lifetimes, low LD and low antibunching dip tells us that red emitting spots are multi-chromophore or multi-chain disordered aggregates that have significant light-harvesting abilities but funnel to multiple low-energy sites within the aggregate. The exact nature of the disorder in the aggregates is a particularly challenging question to address – certainly, these are not ordered or quasi-ordered aggregates such as P3HT[21] as x-ray diffraction studies of PTB7 indicate small fractions of ordered material in solution or film.[22-23] In finding low LD in emission we concur that structural order is low, while the high observed PL brightness indicates that there are potentially many absorbing or emitting chromophores in a single aggregate. We note that the overall solid-state PL quantum yield of PTB7 is very low at ~ 2%.[24] The high overall brightness of the red spots is therefore a consequence of aggregate size, and the large scatter in brightness most likely relates to the variation in size and in effective quantum yield of the single aggregate, i.e. the interaction with localized quenchers. Such variable quenching also explains the large scatter in PL lifetimes of the red species. It would be attractive to be able to correlate the PL intensity of the red aggregates with the size or number of chromophores in the aggregate, but the aforementioned scatter in the PL lifetimes precludes direct correlation. The weak antibunching along with low LD in the red aggregates tells us that there is only a small degree of interaction between chromophores in the aggregate. This is slightly surprising in the context of other photovoltaic polymers (e.g. P3HT), where exciton delocalisation and funneling can enable a large aggregate to behave as a single chromophore.[25-26] We speculate that the disordered nature of the aggregate limits this effect in PTB7.

The most surprising finding is the existence of single-molecule spots that emit at significantly shorter wavelengths than the ensemble optical gap defined by absorption. The nature of these orange and green spots has been examined in some detail in this communication to provide us with more information about the chromophores. First, as a corollary to the discussion above, we propose that the weak emission from the orange and green spots is linked more to the comparatively small amount of absorption presented by a single chain than to any weakening of the transition oscillator strength in comparison to the red spots. The slightly longer PL lifetime and clear mono-exponential decay of these higher-energy chromophores indicates reduced non-radiative losses through dark or charge-pair states[27] compared to the red aggregates. We can explore this a little further by returning to the heated solution experiments that were performed first. There, the PL appeared to get stronger as the solution was heated and red aggregates were broken up into green-emitting single chains. This effect can be quantified by looking at the second-order auto-correlation of the PL from the chains in solution as a function of temperature. This approach is fully detailed in the Supporting Information, along with fits to the correlation functions, and tells us that while the number of emitting particles increases as the temperature of the solution increases, the brightness of them does not change. This observation implies that the green single chains are likely not significantly brighter than the red aggregates, an effect which most likely arises from PL quenching in the aggregates which reduces the PL quantum yield with respect to the isolated chains. The high LD values measured for the orange and green spots offer evidence that the emission chromophores are relatively straight, while the large antibunching dip indicates that we are dealing with a single emitting chromophore rather than multiple emitting species. A schematic of the likely conformational nature of the PTB7 chains for red and green spots is given in the insets of Figure 3d,e.

We now turn our attention to placing these results in context. Three previous experiments have, under specific circumstances, reported PL from energies above the absorption gap in this class of materials[10-12] and attributed the effect to either slow vibrational relaxation enabling emission from a vibrationally 'hot' $S_1$ state, or individual remnant oligomers of the polymer that emit at a shorter wavelength but are present at such low concentrations to be masked in the ensemble absorption. We find that neither of these explanations suitably describes what we see. The hot excited state model, in addition to being rather surprising, should lead to PL from such states having a short lifetime as depopulation to a relaxed $S_1$ as well as to the ground state must occur, yet we find that the lifetime of the green and orange spots is in actual fact *longer* than in the red. The model of remnant oligomers giving rise to the high-energy emission has more merit, but it has been put forward to involve only a single emitting species, consistent with the PL spectrum of an isolated 2mer. We find here, however, that there is in fact a continuum of emitting states, with single-molecule PL peaks ranging from 600 nm through to the red ensemble PL peak at 750 nm, indicative of aggregation leading to the red shift. Furthermore, compelling evidence for aggregation arises from the PL of the heated solution, where continuously blue-shifting PL spectra were observed with increasing temperatures, something that would not be consistent with a sample containing static aggregates and remnant oligomers. In addition, the previously reported[10] effective blue shift of the absorption spectra in a very similar polymer in heated solutions gives further confidence in this conclusion. It has to be conceded that what we detect in this work is PL, so we have no way to comment directly on what the absorption spectra are of the species reported here, but it is reasonable to assume that they would approximately be a high-energy mirror image of the PL. The spot with an $R_{RG}$ of -0.3 and a calculated PL peak of 600 nm should then, with a Stokes' shift similar to that found in the ensemble, have an absorption edge of ~560 nm.

Finally, some context should be provided for what our observations imply in terms of how PTB7 functions in organic photovoltaic cells and how the apparent versus actual optical gap alters the understanding of what may be important in materials used in them. Great efforts have been expended in designing photovoltaic materials with specific optical and electronic properties.[28-29] Often, this can involve trial and error iterative testing in order to find the correct combination of properties. Materials that may appear to be exactly what is required on paper then perform very poorly in devices, or presumed poor materials turn out to have excellent performance. Indeed, it can be speculated that many high-efficiency materials may already exist, but the correct processing or combinations of different materials to achieve that high efficiency may not be known, and thus their true performance remains unharnessed. To that end, the fact that the true optical gap of PTB7, i.e. unaffected by aggregation, is significantly *higher* in energy than was presumed is important to note. Not only does this realization alter how one views the material but it also implies that strong red absorption (essentially all absorption >600 nm) in this material is entirely due to intra- and interchain transitions in the disordered aggregates. The PL from these states is weak, and the lifetime is fast, consistent with disordered aggregation and fluorescence quenching on defects or charges, but that does not necessarily matter in the context of photovoltaic performance, where the aim is to generate charge pairs rather than PL. We therefore speculate as to what other common materials, processed to produce amorphous films, could in fact be processed in a manner that creates aggregates that have desirable photophysical properties – are such situations more common than is often believed, or does PTB7 belong to a special category?

To conclude, we have investigated the high-performance low-gap polymer PTB7 with single-molecule spectroscopy. We find that the commonly observed red PL actually arises from aggregated chains that have unpolarized emission with a short lifetime and show almost no photon antibunching, consistent with the involvement of multiple emitting chromophores. At single-molecule concentrations, however, non-aggregated chains are found to exist which have emission at significantly higher energies (~0.4 eV) than the red aggregates, with – surprisingly – their PL spectra lying entirely within the ensemble absorption spectrum. These non-aggregated chains show strongly polarized PL consistent with straight chromophores which have longer lifetimes along with clear photon antibunching, indicative of a single emitting chromophore being the origin of emission.

**Experimental Methods**

*Materials.* Poly([4,8-bis[(2-ethylhexyl)oxy]benzo[1,2-b:4,5-b']dithiophene-2,6-diyl][3-fluoro-2-[(2-ethylhexyl)carbonyl]thieno[3,4-b]thiophenediyl]) – henceforth called **PTB7**, poly(methyl-methacrylate) (**PMMA**), chlorobenzene and *ortho*-dichlorobenzene were all purchased from Sigma Aldrich Co. and used without further purification.

*Sample Preparation.* For single-molecule experiments PTB7 was dissolved in chlorobenzene and heated at 50°C on a hotplate for ~ 30 minutes. A separate solution of PMMA ($M_W$ = 97 kDa, $M_n$=46 kDa) was made at a solute-to-solvent concentration of 1%, such that films of ~50 nm thickness are created when spinning at 2000 rpm on a spin coater. Doping of PTB7 in the PMMA film at single-molecule concentrations were created by engaging in series dilutions of the PTB7 solution before mixing with the PMMA solution and then spin coating on cleaned 130 μm thick coverslips until small diffraction-limited spots were observed on the microscope. The sample was measured under sealed nitrogen flow to exclude oxygen during measurements.

*Single Molecule Measurements.* Single-molecule experiments are performed on a homebuilt confocal scanning fluorescence setup that has an Olympus IX71 as its base. The 488 nm excitation light was from a Picoquant laser diode (PicoQuant, LDH-D-C-485) which was passed through a clean-up filter (AHF Analysentechnik, z485/10) before being expanded and collimated to ~ 1 cm diameter prior to being coupled into an oil-immersion objective lens (Olympus, UPLSAPO 60XO, NA=1.35). A dichroic filter was used inside the microscope (AHF Analysentechnik, z488RDC) to separate the PL from the excitation inside the microscope. A further clean-up long-pass filter was used on the PL exit port to remove any residual laser excitation before the PL was passed on to the different detector options as shown in the main text Figure 2. The sample was moved with a PI piezo stage (P-527.3CL) controlled along with data acquisition by a Labview program. PL spectra were recorded with an Andor iDUS DU401A-BV CCD coupled to an Andor SR-303i-B spectrograph. Temporal PL dynamics were recorded with pairs of avalanche photodiodes, with the $R_{RG}$ detected on two PicoQuant, τ-SPAD-20 APDs while the linear dichroism and photon antibunching experiments used two Laser Components COUNT-T100 APDs, with free-space asymmetric focussing optics used to ensure APD afterglowing would not be measured in the antibunching experiments. The 646 nm dichroic used to split the PL for creation of the $R_{RG}$ value was an AHF Analysentechnik F33-646. Time tagging of photons was achieved with a PicoQuant, HydraHarp 400 TCSPC system and all data were analysed with Matlab scripts.

**Supporting Information**

Temperature dependent PL spectra & FCS measurements, individual spot traces, the methodology for the calculation of PL spectra from the recorded $R_{RG}$ values, and the concentration independence of $R_{RG}$ are all provided in supporting information.

**Acknowledgements**

We are indebted to the European Research Council for support through the Starting Grant MolMesON (No. 305020).

# Determining the True Optical Gap in a High-Performance Organic Photovoltaic Polymer Using Single-Molecule Spectroscopy

Gordon J. Hedley*, Florian Steiner, Jan Vogelsang and John M. Lupton

Institut für Experimentelle und Angewandte Physik, Universität Regensburg, D-93040, Regensburg, Germany

# Supporting Information

Email: Gordon.Hedley@physic.uni-regensburg.de

**Temperature-Dependent PL Spectra & FCS**

PTB7 solutions dissolved in *ortho*-dichlorobenzene were heated using a homebuilt brass sample chamber placed on the confocal microscope with a long working distance objective. Heating and cooling was provided by resistive heating elements embedded in the sample chamber block and by Peltier elements used for fine control and subsequent cooling. Solution temperatures were monitored, stabilised and recorded with a Pt200 sensor mounted on the outside of the block. PL spectra were recorded after ~5 minutes stabilisation time at each temperature, and the return 20°C spectra was taken after ~1 hour of cooling time with cooling assisted by the Peltier elements, as shown in Figure S1. We note the residual blue-shift and reduced red intensity that remains on the cooled spectrum, indicating that such accelerated cooling of the solution ensures that full aggregation does not automatically and quickly return.

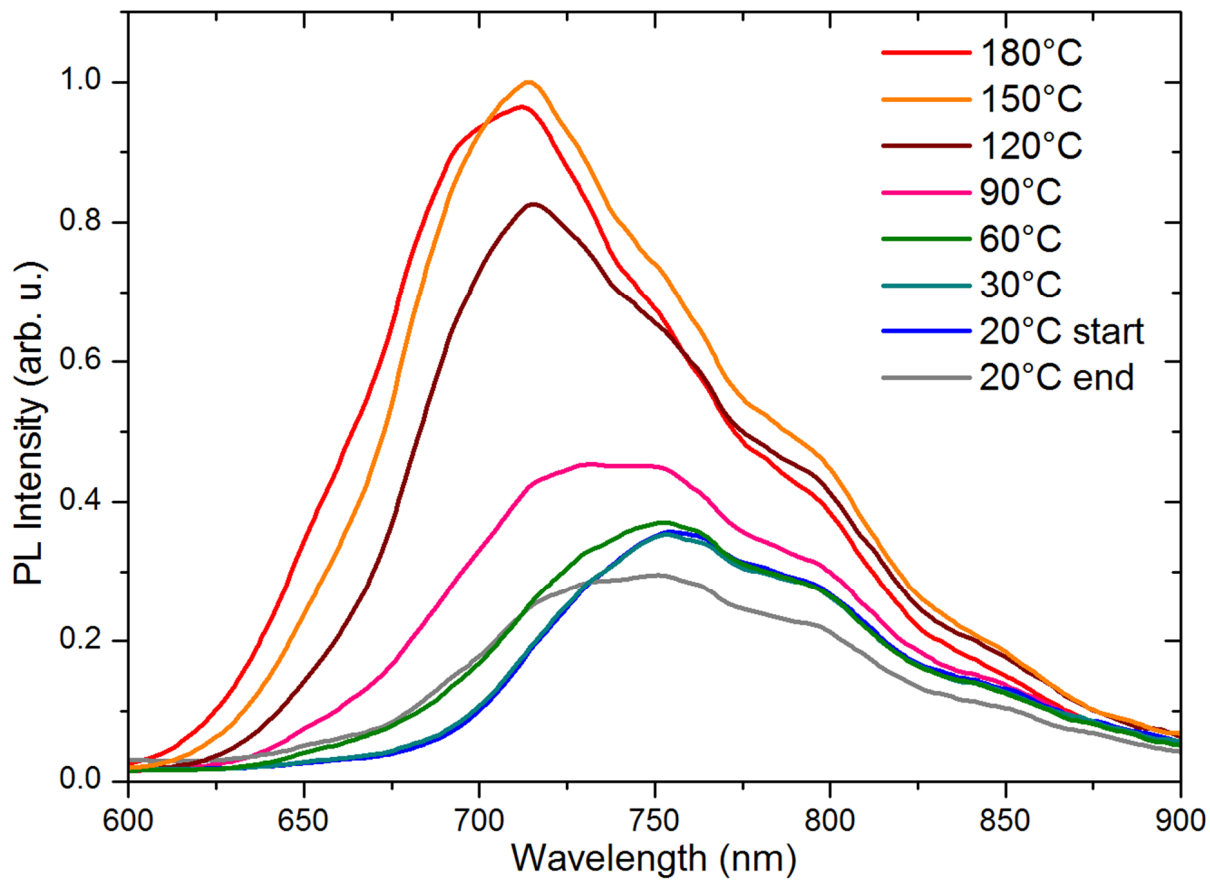

**Figure S1:** PL spectra of PTB7 in *ortho*-dichlorobenzene at denoted temperatures. The y-scaling is global, i.e. comparison of PL intensities at different temperatures can be made directly from this plot.

Fluorescence correlation measurements were performed simultaneously by directing 30% of the PL from the microscope to an APD and recording the time-tagged photon stream. Second order fluorescence correlations were calculated according to equation S1:

$$g^{(2)}(\Delta\tau) = \frac{\langle I(t)I(t+\Delta\tau)\rangle}{\langle I(t)\rangle\langle I(t+\Delta\tau)\rangle} \tag{S1}$$

where the angular brackets denote averaging and I is the PL intensity at a time t and at a time t+Δτ later. The autocorrelation was calculated and is plotted from Δτ=20+ μs to ensure APD afterpulsing is eliminated for each temperature, and is shown in Figures S2-4.

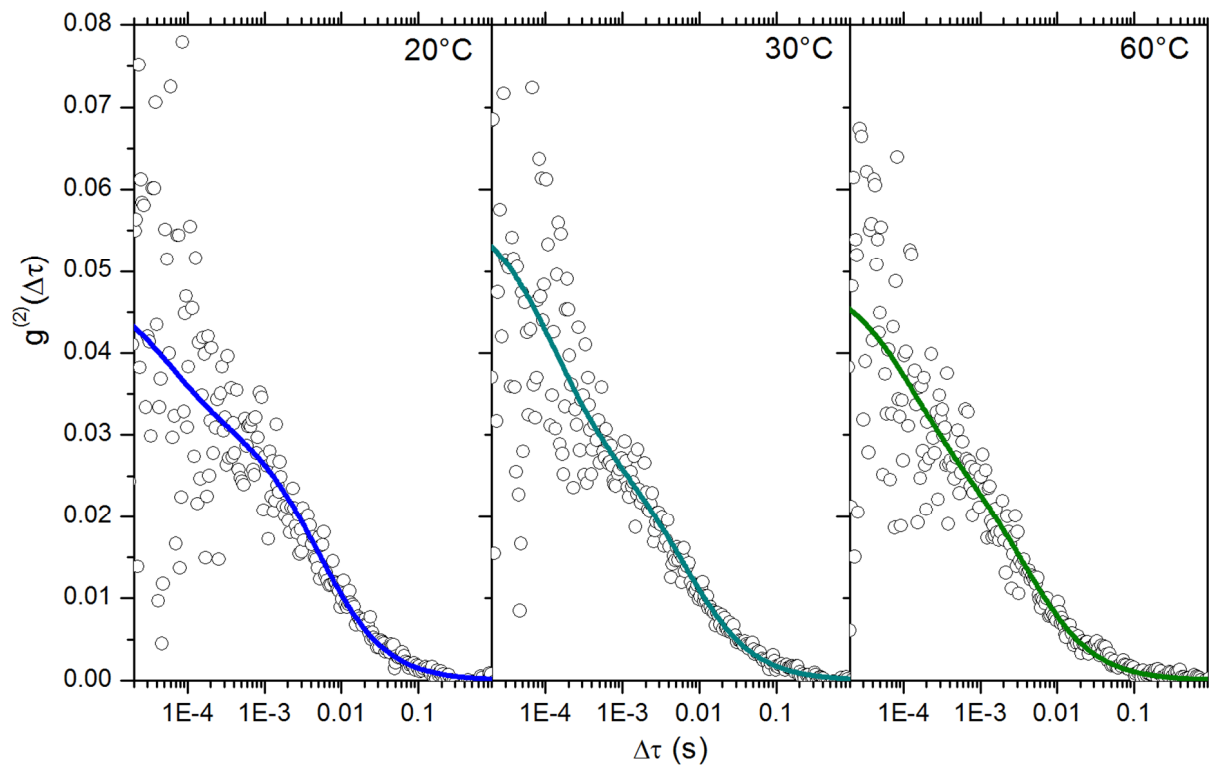

**Figure S2:** Second order auto-correlation plots for PTB7 at very low concentration in solution at the denoted temperatures, also shown are fits as described in the text below.

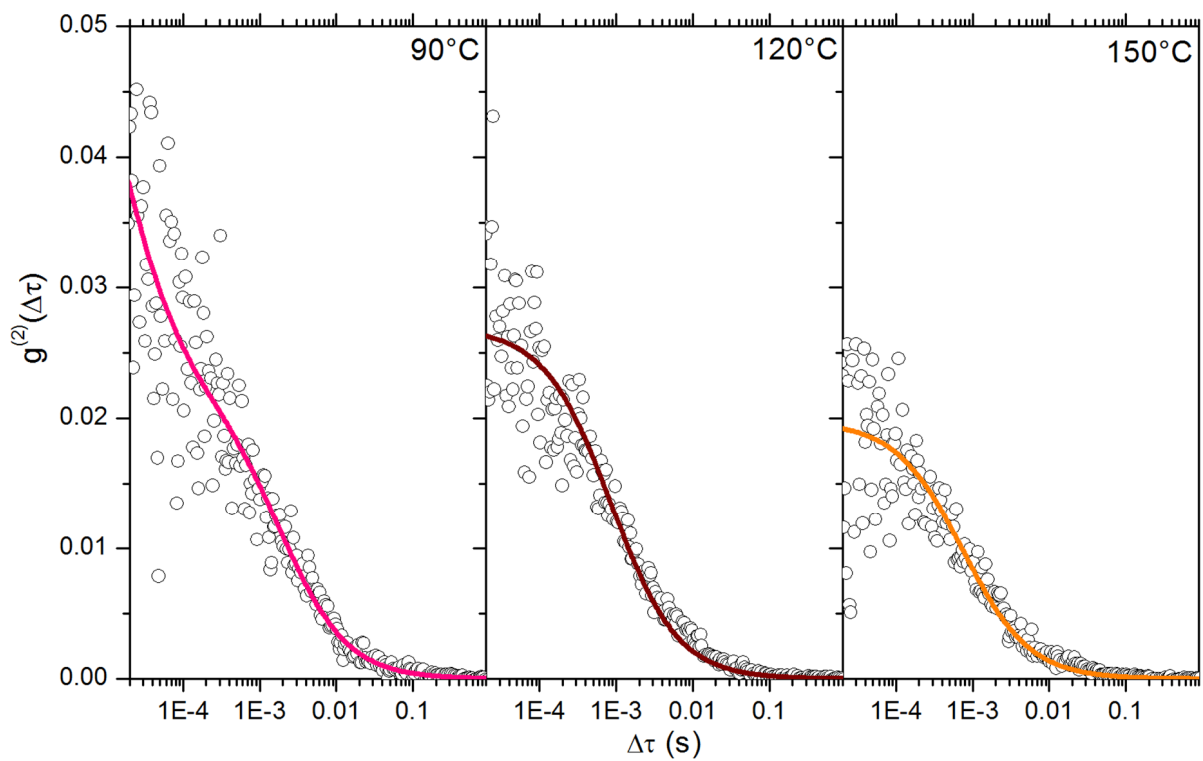

**Figure S3:** Second order auto-correlation plots for PTB7 at very low concentration in solution at the denoted temperatures, also shown are fits as described in the text below.

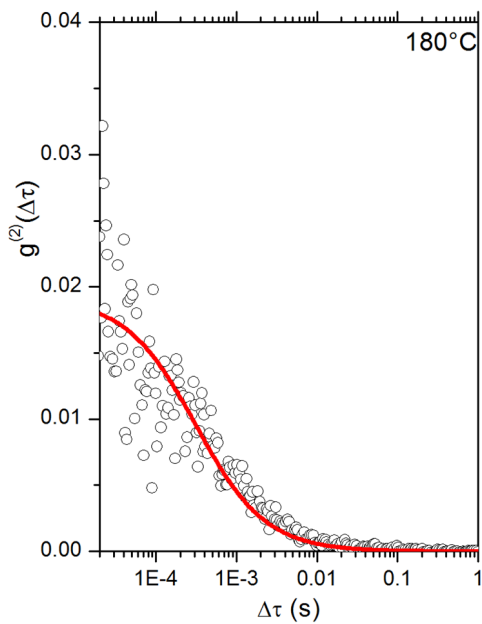

**Figure S4:** Second order auto-correlation plot for PTB7 at very low concentration in solution at 180°C, also shown is the fit as described in the text below.

The solid coloured lines on each FCS trace are fits to 2-dimensional pure diffusion, according to equation S2:

$$g^{(2)}(\Delta\tau) = g(0)_1 \frac{1}{1+\frac{\Delta\tau}{\tau_{diff1}}} + g(0)_2 \frac{1}{1+\frac{\Delta\tau}{\tau_{diff2}}} \quad (S2)$$

Where Δτ is the time, and the fit parameters are $\tau_{diff}$ (time constant of diffusion) and G(0) the amplitude for each component. Two components were required for the 20-90°C traces, and only 1 required for the 120-180°C range. For two components, g(0) = g(0)$_1$ + g(0)$_2$.

g(0) is related to the number of emitting particles, N, traversing the focal volume via equation S3:

$$N = \frac{1}{g(0)} \tag{S3}$$

and thus from the fits we can determine the number of emitting particles as a function of solution temperature, as shown in Figure S5, with errors derived from standard deviations of the fit error.

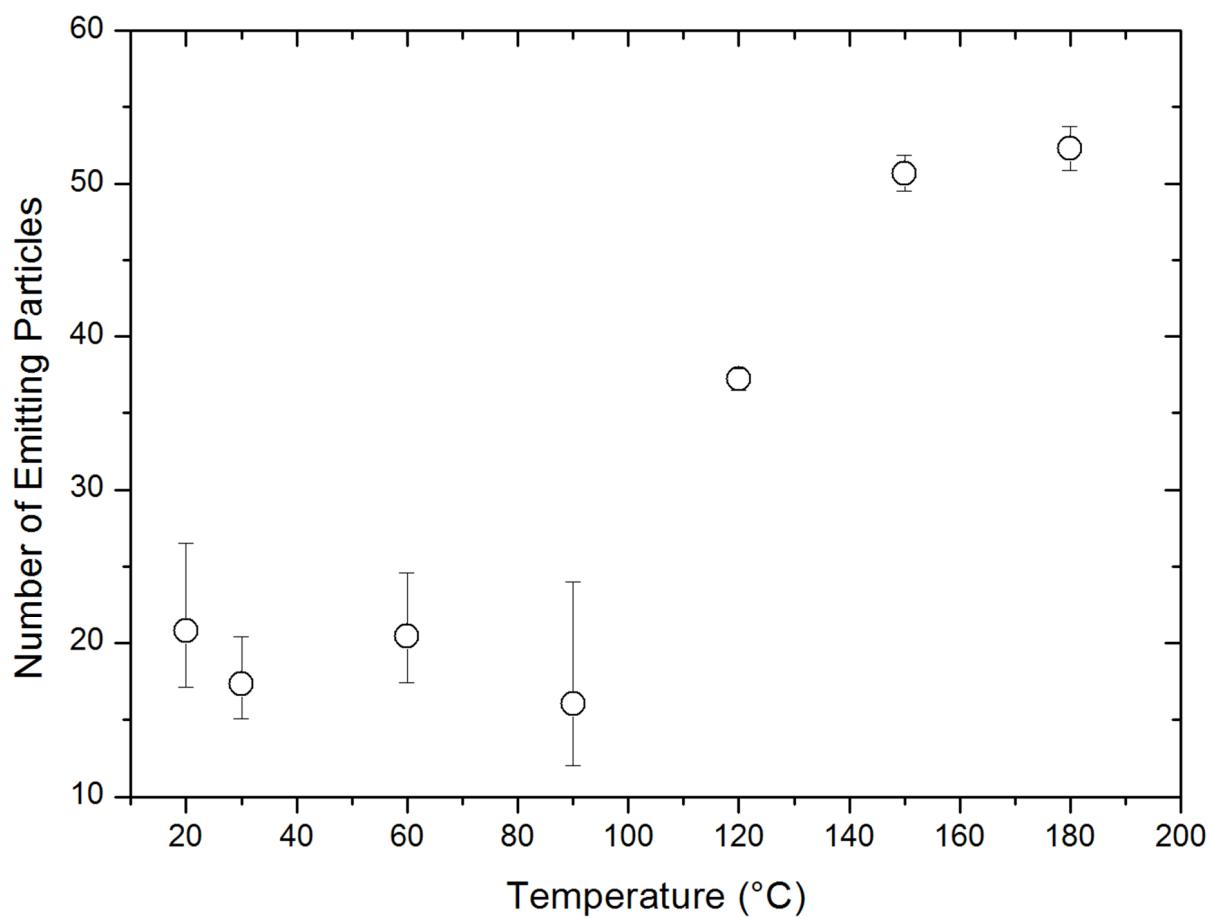

**Figure S5:** Number of emitting particles at each temperature of the solution, as determined from the fits of the measured second order auto-correlation. Error bars are derived from the standard deviations of the fits.

It is clear that the number of emitting particles increases substantially (by a factor of ~ 2.5) as one goes from room temperature to 180°C, consistent with breaking up red aggregates into a larger number of smaller green single chains, and in agreement with the measured PL spectra.

Since we also measure the overall PL intensity with the APD, we can divide this by the number of emitting particles to get the particle brightness as a function of temperature, as shown in Figure S6.

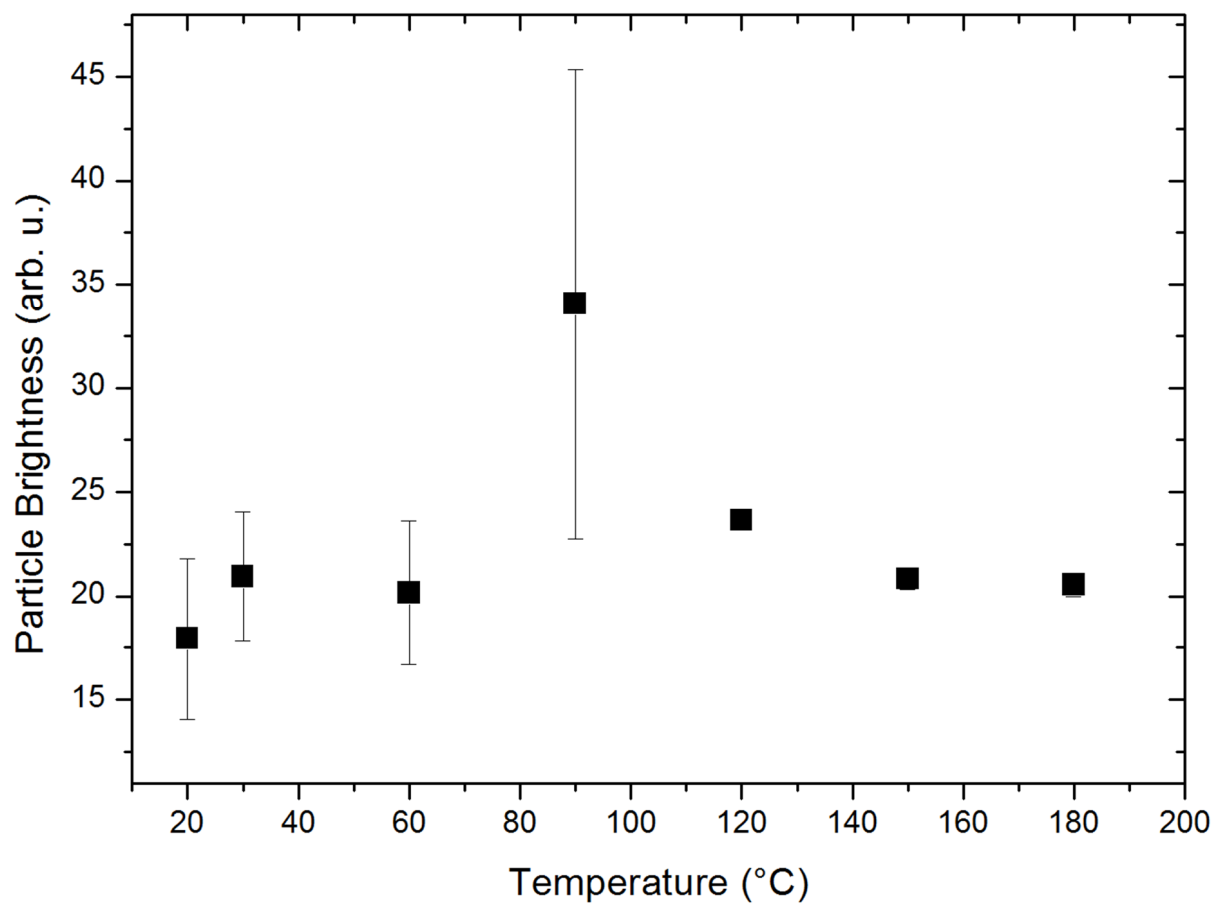

**Figure S6:** The calculated brightness of each particle at each temperature in the solutions of PTB7. Error bars are derived from the errors on the number of emitting particles.

Apart from a single outlier point (with a large associated error), the emitter brightness's do not change much as a function of solution temperature.

**Individual Spot Traces**

Single diffraction-limited spots were identified by recording linescan images with the APDs as described in the main text. Individual example traces for red, orange and green spots are shown below in Figure S7, plotting the total PL intensity (a), the $R_{RG}$ ratio (b), the linear dichroism (c) and the extracted PL lifetimes (d).

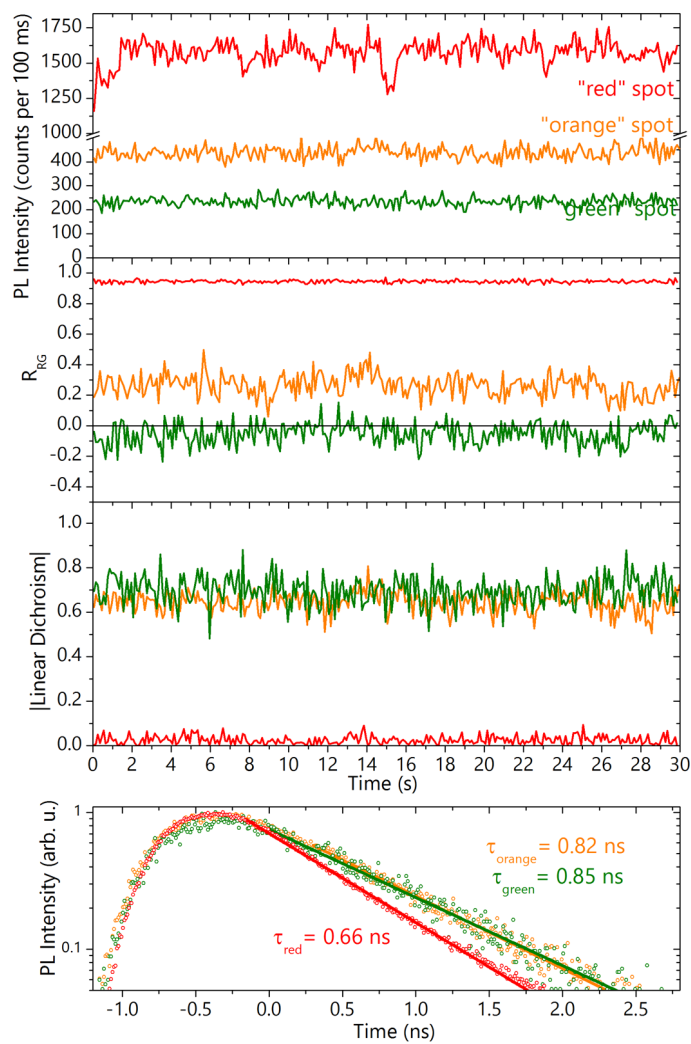

**Figure S7:** Dynamical measurements of single spots. Shown in **a)** is the total PL intensity over 30 seconds with a 100 ms binning for three spots coded with the same colors as in Figure 1 and 2 of the main text. Shown in **b)** is the $R_{RG}$ dynamics of those three same spots, while in **c)** the linear dichroism measured at the same time is shown. In **d)** the PL decays of the three spots are plotted, with the red spot decaying fastest, and the orange and green measurably slower, with the fitted mono-exponential time constants as noted in the inset.

**Simulation of PL Spectrum from Recorded $R_{RG}$ Value**

Full PL spectra of some spots were measured using a spectrograph-coupled CCD, allowing us to record the red, orange and green spectra shown in the main text. However, the weak PL from green spots, combined with photobleaching, led us to have limited opportunities to explore just how short a wavelength such spectra could be.

Thus to enable large numbers of spots to be measured, the red-green fraction, $R_{RG}$, was measured by using a 646 nm dichroic beam splitter to separate PL onto two APDs, as shown in Figure 2 of the main text, with $R_{RG}$ defined as:

$$R_{RG} = \frac{I_{Red} - I_{Green}}{I_{Red} + I_{Green}} \tag{S4}$$

This approach gave us a large number (~1000) of $R_{RG}$ values, with most clustering around $R_{RG}$ ~ 1, representing red aggregated PTB7 chains. Straighter, isolated chains were found to have $R_{RG}$ values in the range -0.3 to 0.85. It was thus desirable to be able to approximately convert the measured $R_{RG}$ values into likely original spectra. We note that this cannot be an exact operation as, fundamentally, information is lost when using the dichroic beam splitter, but by making the assumption that the shortest wavelength PL spectrum that we actually measure (green spectrum, Figure 1 and 3 in the main text) is representative of PL spectra in this emission region, we can make a reasonable estimate of the PL spectrum corresponding to the bluest $R_{RG}$ value.

The main variables that have to be taken into account is the transmission/reflection of the 646 nm dichroic beam splitter, the transmission of the 50:50 beam splitter (the two $R_{RG}$

APDs only receive half the PL, the other half goes to the linear dichroism APDs as shown in Figure 2 of the main text) and the wavelength dependence on the quantum efficiency of the two APDs. These variables were obtained from manufacturer datasheets for each of the components and are shown below in Figure S8.

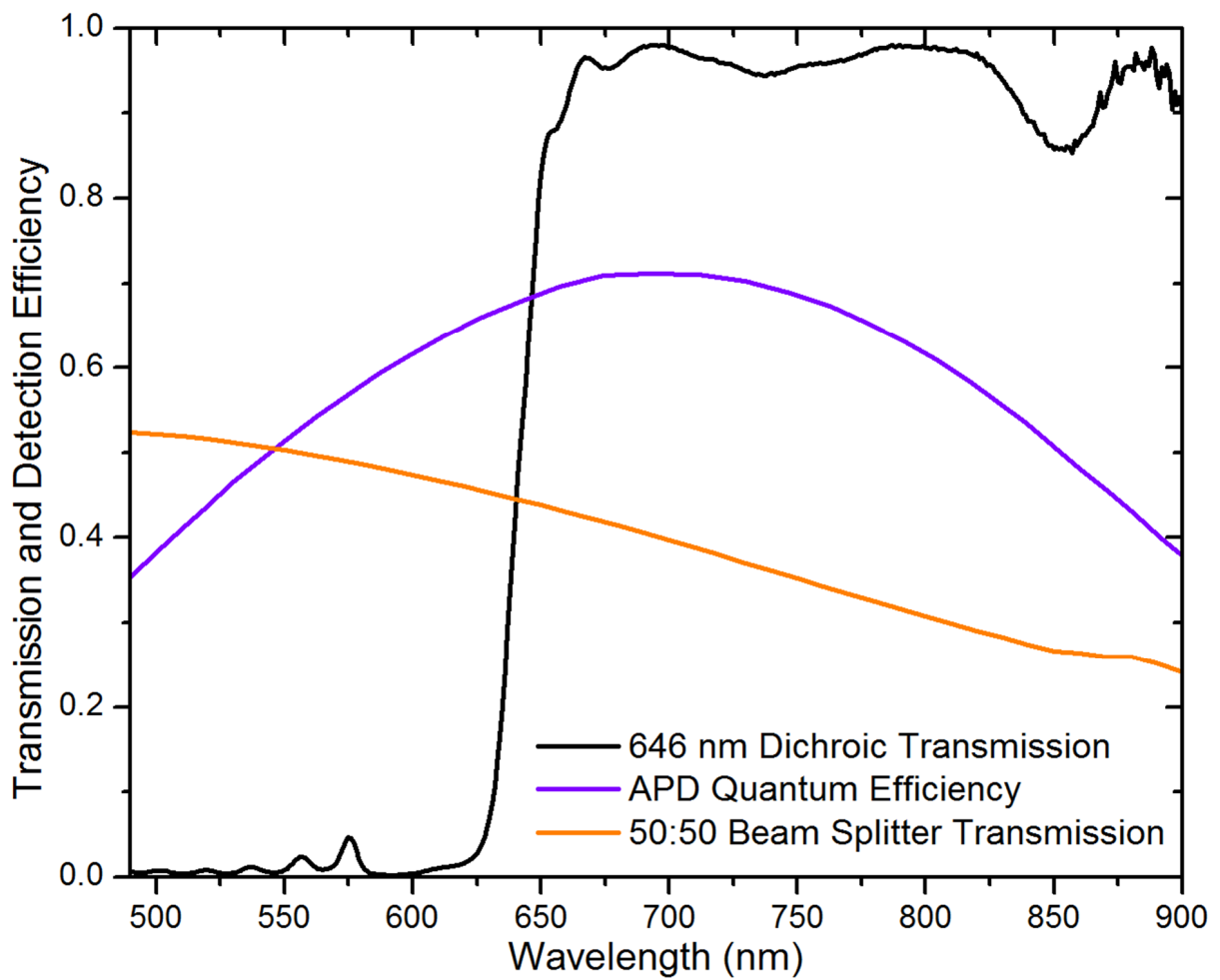

**Figure S8:** Relative efficiency of each element used in the detection of the $R_{RG}$ ratio for the red APD.

Simple multiplication of each of these curves thus gives the final wavelength-dependent sensitivity on each APD (the dichroic transmission is used for the red APD and the reverse for the green APD), as shown below in Figure S9.

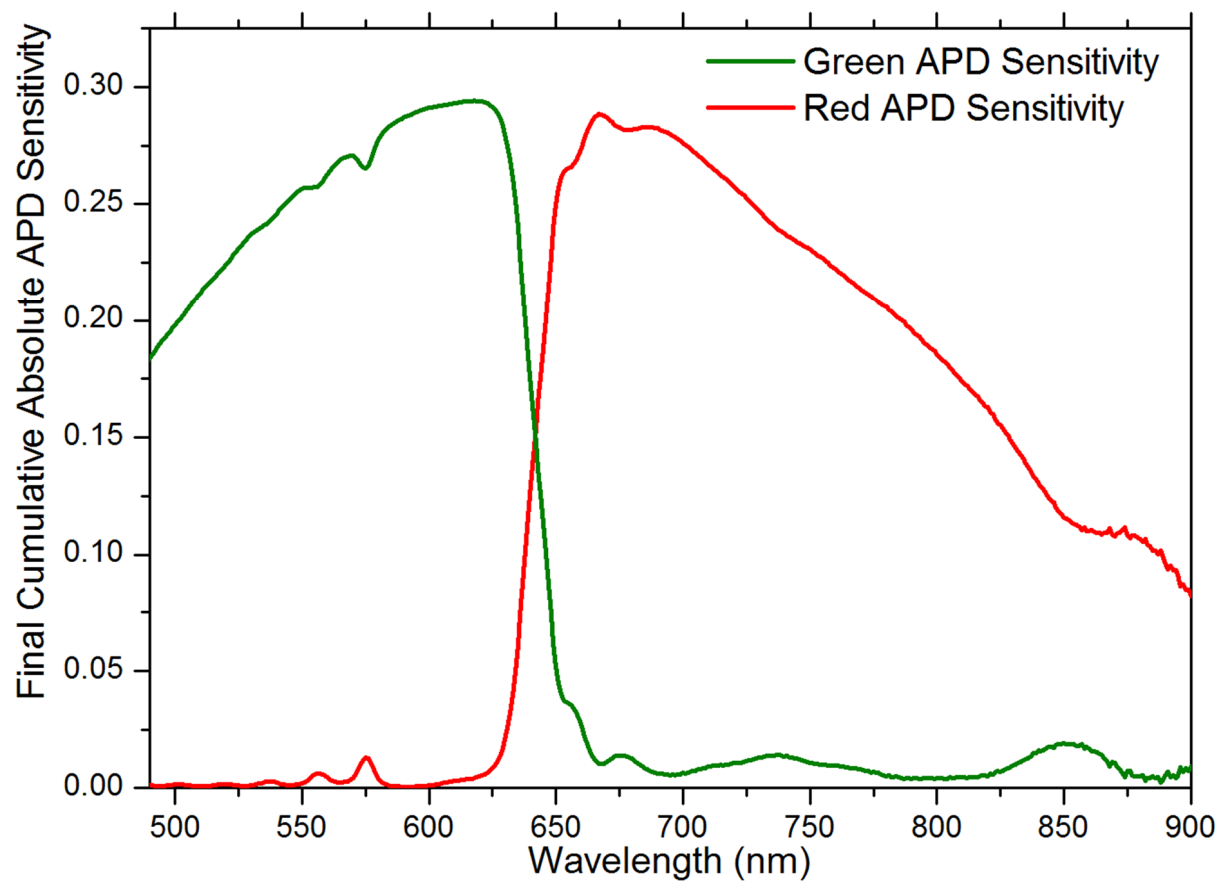

**Figure S9:** Final cumulative APD sensitivity for the red and green APDs when incoming PL is split by the 646 nm dichroic.

By taking an observed PL spectrum and multiplying it by the two final APD sensitivity curves we can obtain the $I_{Red}$ and $I_{Green}$ values that would be recorded on both APDs, and using Equation S4 we can thus calculate the $R_{RG}$ value. This approach is carried out for the measured spectra in Figure 3 and the arrows pointing to the $R_{RG}$ axis above in that figure indicate the calculated values.

In order to extrapolate what the PL spectrum would be for the lowest values of $R_{RG}$ observed (i.e. -0.3) we take the measured green PL spectrum, convert it to the energy scale, fit it with a sum of Gaussians to properly reconstruct the red tail and then translate this spectrum along the energy axis. We convert the spectrum back to wavelengths at various translation points and use that to calculate the $R_{RG}$ value, with the blue dotted line in Figure 3 of the main text showing the PL spectrum that gives an $R_{RG}$ of -0.3.

**Concentration Independence of $R_{RG}$**

To test the indivisibility of the chains that we distribute at very low concentration in a PMMA matrix, we have performed a concentration series to monitor the number of green and red pixels as a function of the concentration. From the master solution dilutions of $2\times10^{-2}$, $2\times10^{-3}$ and $1\times10^{-3}$ were made and films spun. 40x40 μm PL scan images of these three concentrations are shown in Figure S10-12, with the spot density approximately following the concentration.

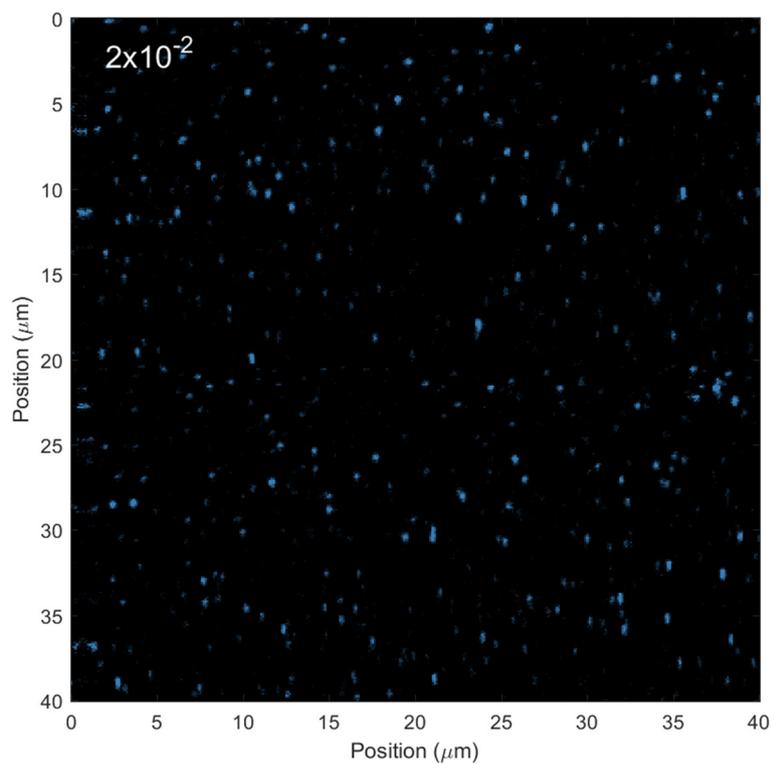

**Figure S10:** PL scan image of a film of PTB7 doped in PMMA at a dilution of $2\times10^{-2}$ from the master solution.

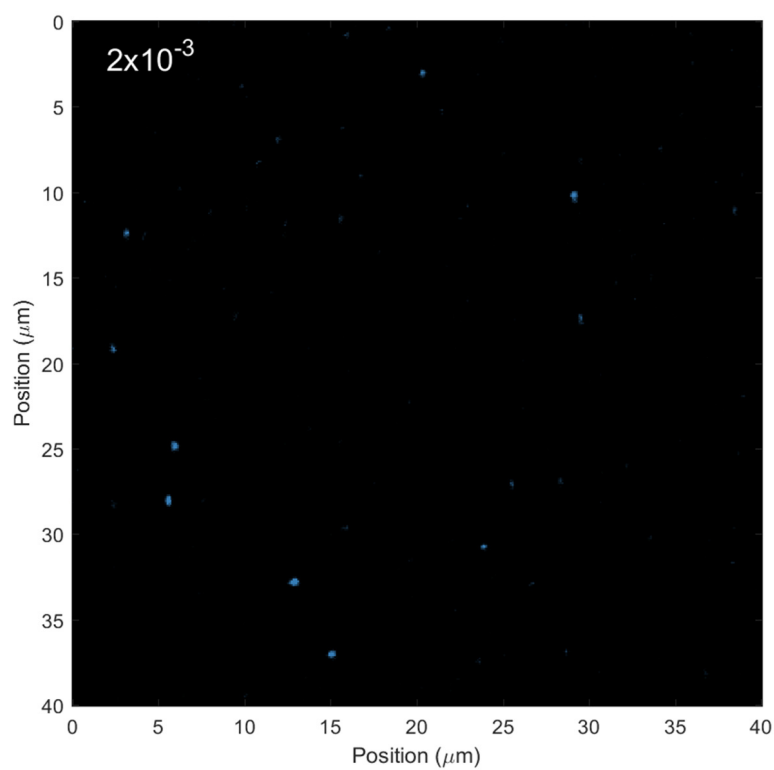

**Figure S10:** PL scan image of a film of PTB7 doped in PMMA at a dilution of $2\times10^{-3}$ from the master solution.

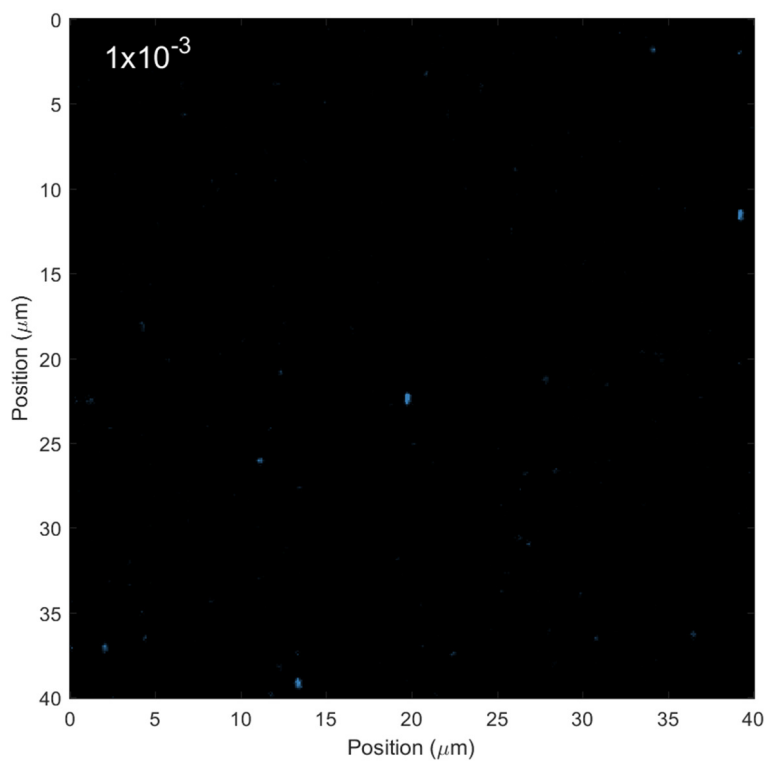

**Figure S10:** PL scan image of a film of PTB7 doped in PMMA at a dilution of $1\times10^{-3}$ from the master solution.

In order to fairly asses the effect of the concentration on the $R_{RG}$ values we have chosen to count *pixels* in the images that are above a noise floor threshold. This is done by recoding the scan images on two APDs with the 646 nm dichroic beam splitter as described in the main text experimental methods. Noise floors of 2 and 7 counts per 2 ms are applied on the green and red APDs, respectively. This sets pixels below these count values to 0 so as to ensure that the $R_{RG}$ map does not contain spurious pixels derived from noise. The $R_{RG}$ ratio is then calculated according to equation S4 on the images, and pixels binned with 0.1 step of $R_{RG}$ are counted across two separate 40x40 μm images and the numbers averaged and normalised to percentages. These percentages for the three concentrations are shown in Figure S13, and indicate that the percentage/distribution of $R_{RG}$ values does not alter appreciably across the factor of 10/20 concentration that is measured here.

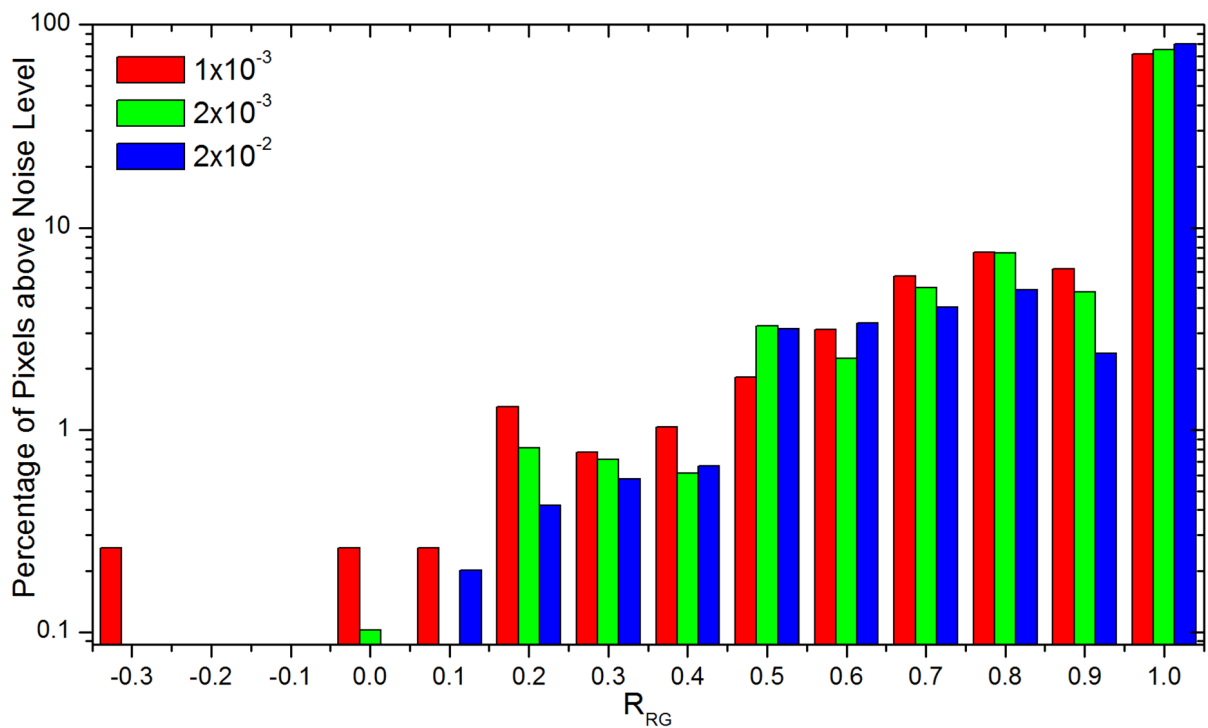

**Figure S13:** Binned distribution of $R_{RG}$ pixel percentages across the three concentrations of PTB7 doped in PMMA. The distribution does not change applicably across this concentration range. Note the y-axis is logarithmic.